
\input phyzzx.tex
\input phyzzx.fonts
\input psfig

\titlepage
\hskip 5.4in Crete-94-12

\title{VORTEX MOTION IN CHARGED FLUIDS}

\author{ { G.N. Stratopoulos }    \foot{
email: stratos@iesl.forth.gr
}
\vskip 0.4cm
\centerline {and}
\vskip 0.4cm
\centerline{ { T.N. Tomaras}  \foot{
email: tomaras@iesl.forth.gr  } } }
 \address{Physics Department, University of Crete and \break
 Research Center of Crete,714 09 Heraklion, Crete, Greece }
 \vskip 1.5cm
\centerline{\bf abstract}

A non-relativistic scalar field coupled minimally to electromagnetism
supports in the presence of a homogeneous background electric charge
density the existence of smooth, finite-energy topologically stable
 flux vortices. The static properties of such vortices
are studied numerically in the context of a two parameter
 model describing this system as a special case. It is shown that
the electrostatic and the mexican hat potential terms of the energy
are each enough to ensure the existence of vortex solutions. The
interaction  potential of two minimal vortices is obtained for
various values of the parameters.
It is proven analytically that a free
isolated vortex with topological charge $N\ne 0$ is spontaneously
pinned, while in the presence of an external
force it moves at a calculable speed and in a direction
$(N/|N|)\;90^0$ relative to it. In a homogeneous external
current $\tilde {\bf J}$ the vortex velocity is $\bf V=-\tilde J$.
Other theories with the same vortex behaviour are briefly discussed.

  \vfill\eject

 \chapter{Introduction}
\REF\rPTb{N. Papanicolaou and T.N. Tomaras, {\it Phys. Lett.}
 {\bf 179A} (1993) 33 }
\REF\rGP{L. Landau and E.M. Lifshitz, {\it Statistical Physics},
Part 2.  Pergamon Press 1980}
\REF\rDo{A.T. Dorsey, {\it Phys.Rev.} {\bf B46} (1992) 8376, and
 references therein.}
\REF\rH{R.P. Huebener, {\it Magnetic Flux Structures in
Superconductors}, Springer 1979}

\REF\rLCF{Qiang Li, J.R. Clem and D.K. Finnemore, {\it Phys.Rev.}
{\bf B43} (1991) 12843}
\REF\rAT{Ping Ao and D.J. Thouless, {\it Phys.Rev.Lett.} {\bf 70}
 (1993) 2158}
\REF\rNO{H.B. Nielsen and P. Olesen, {\it Nucl. Phys.} {\bf 61B}
(1973) 45 }
\REF\rJR{L. Jacobs and C. Rebbi, {\it Phys. Rev.} {\bf B19} (1979)
 4486 }
\REF\rPTa{N. Papanicolaou and T.N. Tomaras, {\it Nucl. Phys.}
{\bf 360B} (1991) 425 }
\REF\rP{N. Papanicolaou, Skew deflection of magnetic vortices in a
field gradient, Crete preprint 1993, to appear in Physica D}
\REF\rC{S. Coleman, {\it Aspects of Symmetry}, Cambridge University
 Press 1988}
\REF\rDe{G.H. Derrick, {\it J. Math. Phys.} {\bf 5} (1964) 1252}
\REF\rA{A.A. Abrikosov, {\it Sov. Phys. JETP } {\bf 5} (1957) 1174}
\REF\rST{G. Stratopoulos and T.N. Tomaras, Crete preprint, to appear.}
\REF\rL{See for instance T.D. Lee, {\it Particle Physics and
Introduction to Field Theory}, Harwood Academic Publishers 1981 }
\REF\rNV{P. Nozi$\grave e$res and W.F. Vinen, {\it Philos. Mag.}
 {\bf 14}  (1966) 667, and references therein.}
\REF\rFGM{J. Friedel, P.G. de Gennes and J. Matricon,
{\it Appl. Phys. Lett.} {\bf 2} (1963) 119}
\REF\rB{G.K. Batchelor, {\it An Introduction to Fluid Dynamics },
Cambridge University Press 1967}
\REF\rLeLe{J-M. L$\acute e$vy-Leblond in {\it Group Theory and its
Applications}, Ed. E.M. Loebl, Academic Press 1971}
\REF\rW{N.R. Werthamer, in {\it Superconductivity} Ed. R.D. Parks,
Marcel Dekker 1969, and references therein.}
\REF\rWZ{X.G. Wen and A. Zee, {\it Phys. Rev. Lett.} {\bf 62}
 (1989) 2873}
\REF\rMBWK{F.G. Mertens, A.R. Bishop, G.M. Wysin and C. Kawabata,
 {\it Phys. Rev.} {\bf B39} (1989) 591}

 \par
We study in detail some aspects of the physics of
flux vortices in systems described by
a non-relativistic complex scalar
field minimally coupled to electromagnetism and in the presence of a
 background
homogeneous electric charge density. The model was introduced and the
essential features of its vortex dynamics were discussed in a recent
publication \refmark\rPTb. Here we elaborate upon and extend
the results of reference [1].

Our motivation to analyze the behaviour of topological solitons in
 the above model is twofold. First, it is actually a whole class
 of models potentially
relevant to the theoretical treatment of several physical systems.
In the absence of any scalar potential
one deals with a plasma with infinitely-massive positive
charges frozen in a configuration
of constant charge density.
With a quartic potential for the scalar
field the theory can be viewed either as the natural coupling
to the electromagnetic field of the dynamical Gross-Pitaevskii
 equation
of superfluidity \refmark\rGP or as one possible dynamical extension
of the static
Ginzburg-Landau model of superconductivity, reminiscent in particular
of the large friction
limit of the system considered in reference [3].
Second, in the context of the above models one can obtain definite
theoretical
predictions about the gross features of the motion of vortices in
two spatial dimensions, without
any approximation and insensitive to
the fine details of the Hamiltonian. This might prove a
reasonable
starting point for the understanding of the dynamics of an isolated
Abrikosov
vortex in thin superconducting films, a subject of considerable
interest in the
context of ordinary as well as high-$T_c$ superconductors
\refmark\rDo \refmark\rH  \refmark\rLCF \refmark\rAT.

In section II we introduce the model \refmark\rPTb.
Since the bulk of our results refer to
two dimensional objects we directly present the model in two spatial
dimensions.
Section III contains our numerical study of axially symmetric
vortex solutions for various
values of the topological charge. We explicitly demonstrate the almost
identical roles of
the mexican hat potential term and the
electrostatic term in the energy density of the model. We show
that either one of these two
terms is enough to
guarantee the existence of non-trivial vortex solutions.
The parameter space is divided into two regions according to whether
the ratio of the energy of the doubly-charged vortex over the energy
of two
well-separated single vortices is smaller or larger than one.
Finally the two-vortex system is discussed and their interaction
potential is
obtained numerically. Contrary to the relativistic model \refmark\rNO
 \refmark\rJR the
interaction energy is not always a monotonous function
of the distance between the two
vortices. In a wide region of the parameter space there is a bump in
the interaction potential of two vortices,
which leads to the possibility of having doubly-charged dumb-bell
shaped static solutions. The dynamics of vortices is discussed in
section IV.
The canonical structure of the model is derived and it is shown that
the
unambiguous linear momentum is essentially the first moment of the
topological density. The momentum
conservation law in the absence of any external fields
is shown to imply the spontaneous pinning of
isolated vortices, while the application of an
external force results in the motion of the vortex at a constant
calculable speed in a
direction $(N/|N|)\;90^0$ relative to it \refmark\rPTb, features
encountered already
in the motion of magnetic bubbles in the
ferromagnetic continuum \refmark\rPTa \refmark\rP.
Finally, in the discusion section V we
show that necessary condition for the manifestation of the above
Hall-behaviour
in the dynamics of topological solitons in a given field theory is
that the
momentum part of its symmetry algebra admits central extension. Other
interesting models
sharing this property are commented upon.

\chapter{ The model - General features}

\par
We will be dealing with a non-relativistic complex scalar field
${\Psi}$ (the condensate) minimally coupled with coupling $q$ to the
electromagnetic
potential $(A_0, A_i)$. To make the model
physically sensible and mathematically
consistent it is necessary
to introduce a background (positive-ion) charge density
to neutralize the system. For simplicity we take it to be constant
and homogeneous throughout. We
concentrate on the physics of infinitely long straight vortices
i.e. on field configurations uniform along the third spatial direction
 and
define the model directly in two space dimensions by the lagrangian
$$ {\cal L}\;=\; {i \gamma\over 2}
[ \Psi \sp\ast {\cal D}_t\Psi \;-\; c.c. ]
 \;+\;\gamma q \Psi_0\sp2 A_0
 -\; {\gamma\sp2\over 2m} \bigl| {\cal D}_i\Psi |^2
+ {1\over 8 \pi} ({\bf E}\sp2 - B\sp2 )\;-\; V(|\Psi|)
  \eqn\etwoone  $$

\noindent
with  $ B= \epsilon_{ij} {\partial_i}A_j $ ,
$ E_i = -{1\over c} \partial_t A_i - \partial_i A_0\;$ ,$\;
{\cal D}_t\Psi=
(\partial
_t+
i q A_0)\Psi \;$,$\; {\cal D}_i\Psi=(\partial_i-i {q\over c} A_i)\Psi
 \;. \;\gamma,\; m$ and $q$ are parameters, c is the speed of light
  and the
 spatial indices $i,\;j$ range from 1 to 2.
As it will become
clear, although the specific form
of the potential $V$ changes the details of the profile of the vortex
solutions,
 it does not affect their dynamical behaviour. For the discussion
 of vortex dynamics $V$ could even be absent but
in order to make contact with a variety of models of physical interest
we will allow for
a mexican hat phenomenological potential
$$ V(|\Psi|)= {1\over 8} \;g\; (\Psi \Psi\sp\ast-\Psi_0\sp2)\sp2
\eqn\etwotwo $$
with quartic self-coupling $g$ .
\noindent
Rescale fields and coordinates according to
$$ x_i\to {\gamma \kappa\over \Psi_0 \sqrt {mg}} \;x_i\hskip 1.4cm
 t\to {\gamma \kappa\sp2\over \Psi_0\sp2 g } \; t  $$
$$ \Psi\to \Psi_0  \Psi \hskip 1.4cm
A_0\to {g \Psi_0\sp2\over \gamma \kappa\sp2 q} \;A_0\hskip 1.4cm
A_i\to {c \Psi_0 \sqrt {mg} \over  \gamma \kappa q} \; A_i
 \eqn\etwothree   $$
to obtain \refmark\rPTb, in terms of the dimensionless quantities $t$,
 $x_i$,
$\Psi$, $A_0$ and $A_i$ used from now-on, the lagrangian
 $$ {\cal L}\;=\;{1\over 2}
( \Psi\sp\ast (i \partial_t - A_0)\Psi \;+\; c.c.)
 \;+\;A_0
   -\; {1\over 2} \;
|D_i\Psi|^2 + {1\over 2} ({1\over \lambda} {\bf
E}\sp2 - B\sp2 )
\;-\; {1\over 8} \kappa\sp2 (\Psi \Psi\sp\ast-1)\sp2
 \eqn\etwofour   $$

\noindent
with $ B= \epsilon_{ij} {\partial_i}A_j $ ,
$ E_i = -\partial_t A_i -\partial_i A_0 $ and $D_i=\partial_i-i A_i$ .
We keep the same symbols
to simplify our notation.

Classically the model depends on the two free dimensionless parameters
 $\kappa$
and $\lambda$
defined by
$$  \kappa\sp2 = {g m\sp2 c\sp2\over 4\pi q\sp2 \gamma\sp4}
 \hskip 1.5cm \lambda= {m\sp3 c\sp4 \over {4 \pi q\sp2 \gamma\sp4\
 \Psi_0\sp2}
  }  \eqn\etwofive $$
The conversion to physical units for the lengths
and the characteristic times is given for the specific system at hand
 by the rescaling formulas \etwothree. An
overall factor $g {\Psi_0}^4 / \kappa^2$ was dropped from
$\cal L$ since it does not enter the equations of motion.
It plays though the role of $1/\hbar$ in the quantum theory
and determines the necessary condition for the validity of our
semiclassical approximation
 $$ {\kappa^2 \over {g {\Psi_0}^4} } \to 0 \eqn\etwosix $$
In this limit we expect the quantum solitons to resemble closely their
classical ascendants studied below.

The equations of motion on the other hand read
$$ i \dot \Psi=-{1\over 2} {{\bf D}\sp
2} \Psi + A_0 \Psi+{1\over 4}
{\kappa}^2 (\Psi\sp\ast  \Psi-1)\Psi  $$
$$ {1\over \lambda}\dot E_i=\epsilon_{ij} \partial_j B - J_i
\eqn\etwoseven $$

\noindent
The Gauss constraint and the remaining Maxwell identity
$${1\over \lambda} \;\partial_i E_i=\rho \hskip 1.5cm \dot B
= -\epsilon_{ij} \partial_i E_j
\eqn\etwoeight  $$
complete the set of field equations of the model. The charge and
current
densities $\rho$ and $J_i$ respectively, are given by
$$ \rho=\Psi\sp\ast \Psi - 1 \hskip 1.5cm J_i={1\over {2i}} [\Psi
\sp\ast D_i\Psi-(D_i\Psi)\sp\ast \Psi] \eqn\etwonine $$
The corresponding energy functional is the sum of four positive terms
    $  W  =  W_d+W_b+W_e+W_v    $ with
 $$ W_d= {1\over 2}\; \int d^2x|D_i \Psi|\sp2
\hskip 1.5cm
 W_b={1\over 2}\; \int d^2x \;{B\sp2}  $$
$$W_e={1\over 2 \lambda}\; \int d^2x \;{{\bf E}\sp2} \hskip 1.5cm
W_v={1\over 8 }  \kappa\sp2 \; \int d^2x\;(\Psi \Psi\sp\ast-1)\sp2
 \eqn\etwoten  $$

We will discuss exclusively smooth, particle-like, localized
configurations
with finite energy. Notice that even in the
absence of the potential $V$ finiteness of $W_e$ already implies that
these configurations must be neutral. A localized configuration with
 non-zero
net electric charge leads to an electric field behaving like $1/r$
at large distances and this makes $W_e$ diverge.
This neutrality requirement translates into a boundary condition for
 $|\Psi|$
at infinity. In fact $|\Psi|$ must tend to one, since otherwise
it will lead to infinite charge at large distances. It is possible
of course to interpolate to a neutral singular configuration but only
at an infinite energy cost. One could in principle contemplate a
logarithmically divergent energy due to a non-zero finite charge but
it is
impossible to allow for an asymptotic value of $|\Psi|$ different
from one.
We conclude that the electrostatic term exactly like
the standard phenomenological
potential term, when combined with the requirement of finite energy,
leads
to the same boundary condition for $|\Psi|$ at infinity. Furthermore,
we will explicitly demonstrate in the
next section that even without $V$ the field equations
do support the existence of non-trivial soliton solutions.
We thus have to
impose the conditions
$$ Q=\int d^2x \;\rho=0 \eqn\etwoeleven $$
and
$$ |\Psi({\bf x})| \rightarrow 1 \hskip .2in and  \hskip .2in
 |D_i\Psi| \rightarrow 0 \hskip .2in as \hskip .2in
       |{\bf x}|
                       \rightarrow
                         \infty   \eqn\etwotwelve $$
Such configurations are known \refmark\rC to be classified
according to
the first homotopy group of $S^1$ into
disjoint topological sectors (equivalence classes)
characterized by an integer topological
charge $N$, computed from
$$ N={1\over 2 \pi i} \int d^2x \;\epsilon_{kl} \partial_k \Psi\sp\ast
\partial_l \Psi \eqn\etwothirteen $$
For field configurations satisfying the above boundary conditions one
may
equivalently use the manifestly gauge-invariant formulas
$$ N={1\over 2 \pi} \int d^2x \;B $$
or
$$N={1\over 2 \pi i} \int d^2x
\;[\epsilon_{kl} (D_k\Psi)\sp\ast (D_l\Psi)-i B (\Psi\sp\ast \Psi-1)]
\eqn\etwofourteen $$
as the integral of the explicitly gauge-invariant topological density
$$\tau={1\over 2 \pi i} [\epsilon_{kl} (D_k\Psi)\sp\ast (D_l\Psi)-i B
(\Psi\sp\ast \Psi-1)] \eqn\etwofifteen $$
An argument which distinguishes these three definitions and favours
the use of the last version will be given
in the next section.

A final technical remark is related to the gauge invariance mentioned
 above.
The action \etwofour\ is invariant under the space dependent gauge
transformations $ {\Psi}^{\prime}=exp{(i \Lambda ({\bf x}))} \Psi$ ,
${A_i}^{\prime}=
A_i+\partial_i \Lambda$, ${A_0}^{\prime}=A_0$ for arbitrary function
$\Lambda ({\bf x})$. For small gauge transformations i.e. for $\Lambda$
such that $\epsilon_{ij}
\partial_i \partial_j \Lambda=0$ the topological charge $N$ is also
invariant.
Large, and for the $U(1)$ case at hand necessarily singular,
transformations for which
$\epsilon_{ij} \partial_i \partial_j \Lambda= 2 \pi n \delta({\bf x})$,
$N$
changes by $n$ units. To solve the field equations one has to eliminate
the
redundant degrees of freedom by imposing a gauge-fixing condition. We
choose to work in the gauge $\nabla \cdot \bf A=0$.

Equations $\etwoseven$ and $\etwoeight$ admit the one parameter
family of trivial vacuum solutions
$$ \Psi=exp{(i \alpha)}  \hskip 1.2cm A_i=0 \hskip 1.2cm A_0=0
\eqn\etwosixteen $$
for arbitrary constant value of $\alpha$.
Although the small oscillation analysis around a trivial vacuum is of
interest by itself and deserves detailed study, the purpose of this
work is
to analyse the physics of flux-vortices to which we now turn our
attention.

         \chapter{ Static properties }

\section{ Isolated vortices}

We will search for
static axially symmetric solutions of the field
equations in any given non-trivial topological sector.
The most general ansatz for such a vortex configuration
with topological charge $N$ is in polar coordinates $(r,\phi)$ given by
     $$ \Psi({\bf x}) \;=\; X(r) \; e\sp  { i N \phi }
     \hskip 1.2cm A_0( {\bf x}) \;=\;f(r) \hskip 1.2cm
    {\bf A}({\bf x}) \;=\;{N \over  r}  (1-\alpha(r)) \hat {\bf \phi}
      \eqn\ethreeone $$

      \noindent
At infinity \etwotwelve\ implies the boundary conditions
    $$ X(r) \rightarrow 1      \hskip .4in \alpha (r) \rightarrow 0
        \eqn\ethreetwo          $$

\noindent
Within the above ansatz the electric and the magnetic fields
are
   $$ {\bf E} \;=\; -f^{\prime} \;{\hat  r} \hskip 1cm
	B\;=\; -{N \alpha^{\prime} \over r} \eqn\ethreethree $$

\noindent
while the electric charge and current densities take the form
   $$ \rho \;=\; X\sp2-1 \hskip 1cm
{\bf J} \;=\; {N\over r} \alpha X\sp2 \hat \phi \eqn\ethreefour  $$

\noindent
Finally the field equations reduce to the following set of ordinary
nonlinear
differential  equations
$$ X'' \;+\;{1\over r}X' \;-\; {N\sp2\over r\sp2} X \alpha\sp2
\;=\; {1\over 2} \kappa\sp2 (X\sp2-1) X \;+\; 2fX $$
$$ {1\over \lambda}\;(f'' \;+\;{1\over r}f') \;=\; - \rho \hskip 2cm
\alpha'' \;-\;{1\over r} \alpha'\;=\; \alpha X\sp2    \eqn\ethreefive $$

The behaviour of the solutions around the origin $r=0$ is dictated by
the
requirement of smoothness and by the equations themselves to be
  $$ X(r) \rightarrow  c_1 \; r\sp{|N|}  \hskip .5in \alpha (r)
  \rightarrow  1 +
    c_2 \; r\sp{2} \hskip .5in f(r) \rightarrow  c_3 + {1\over 4}
    \lambda \;
    r\sp{2}
        \eqn\ethreesix          $$

\noindent
where $c_1$, $c_2$ and $c_3$ are undetermined constants.

Being unable to solve analytically the equations at hand we proceeded
numerically.After
some experimentation we decided that among the various methods available
the variational approach is the most convenient.
The method is based
on the fact that the field
equations
$\etwoseven\,$ and $\etwoeight\,$ for static solutions are identical
to the conditions for the minimization of the energy functional
\etwoten\ under the Gauss-law and charge neutrality constraints.
This is true even after we insert into the energy functional the
spherically symmetric ansatz \ethreeone\ \refmark\rC and leads in that
case
directly to equations $\ethreefive$.
We can thus use the variational method and minimize the
energy in order
to find approximate
solutions of our equations.

We approximate the fields of the ansatz
by  a set of trial functions depending on some number of variational
parameters. We then evaluate the energy as a function of these parameters
 and
minimize the resulting expression. The position of the minimum in the
parameter space determines the approximate solution, whose energy
is the value
of the energy functional at the minimum.
  The variational ansatz we used, compatible with the boundary
 conditions at large $r$  is :

	$$ X(r)\;=\; 1\; +\; e\sp{-k r} \;
\sum_{l=0}\sp{m} z_l {r\sp{l}\over l!} \hskip 1.5cm
	 \alpha (r)\;=\;  e\sp{- r} \;
\sum_{l=0}\sp{m} \alpha _l {r\sp{l}\over l!} \eqn\ethreeseven $$

	$$ f'(r)\;=\; e\sp{-k r}\; \bigl[
{c_{-1}\over r}\; +\; \sum_{l=0}\sp{m} c_l {r\sp{l}\over l!} \bigr]
        \; +\; e\sp{-2 k r}\; \bigl[
{d_{-1}\over r}\; +\; \sum_{l=0}\sp{2m} d_l {r\sp{l}\over l!} \bigr]
 \eqn\ethreeeight $$
The asymptotic analysis of $\ethreefive$ requires that $k=\kappa$
in the above equations. Numerically though it is more accurate
to leave $k$ as an independent variational parameter. The optimum
value of $k$ is not $\kappa$. This is no contradiction. The asymptotic
study actually refers to very large distances, where all fields
have already reached their vacuum values and do not
contribute
to the energy. One may in principle put $k=\kappa$
but this frustrates the other variational parameters, which are
forced to take unaturally diverse values in order to lead to
the right solution. This induces in general higher numerical errors.

Plugging the above expressions for $X$ and $f^\prime$ into Gauss'
constraint
equation (2.8a) we determine the coefficients $c_{-1}$, $d_{-1}$,
 $c_i$ and
$d_i$ in terms of the $z_i$. The boundary conditions at the origin
give $\alpha_0\;=\;\alpha_1\;=\;1,\; z_0\;=\;-1$
for the $N=1$ vortex. For higher-$N$
vortices one has more conditions due to the faster fall-off of $X$
at $r=0$.
Thus, for the $N=2$ vortex we obtain in addition to the above that
$z_1=
-k$. Finally, the vanishing of the total electric charge eliminates
one more of the unknown coefficients of the variational ansatz. In our
computations as it will be explained below, we achieve satisfactory
accuracy
for $m = 10$, i.e. with 16 and 15 independent variational parameters
for $N=1$
and $N=2$ respectively.

Having solved explicitly Gauss' constraint as well as the neutrality
condition, we insert the variational ansatz into the energy functional,
we compute analytically the spatial integrals to
end-up with a polynomial of fourth order in the variational
parameters. A quasi-Newton minimization procedure in then used to
evaluate
its minimum. The method converges fairly rapidly within four or five
iterations
inspite of the large number of parameters.
The physical nature of the problem and the proper choice of the
variational
ansatz which restricts the fields to the right subspace of the
configuration
space, are the two factors to which it is reasonable to attribute
this rapid convergence.

Figure 1 shows the profiles of the $N=1$ vortex solutions for two
different sets of values of the parameters of the model.
In figures 1a and 1b
we plot the
magnitude of the electric current $J$,
the magnetic field $B$,
the charge density $\rho$ and the magnitude of the electric field $E$
devided by $\lambda$ as
functions of the radius $r$ from the center of the vortex for
$\lambda=0.1$ and $\kappa=2$. Similarly, in figures 1c and 1d
we have the same quantities but for $\lambda=0.1$ and $\kappa=0$,
the case
with no quartic potential.

Before we proceed with further results a few comments about the
accuracy of our numerical computations are in order. To obtain an
estimate of the
error induced by the truncation of the configuration space, we minimized
the energy for various values of $m$ in order to check
the stability of our results against a broadening of the trial
ansatz $\ethreeseven\;$, $\ethreeeight\;$.
The energy at the minimum for $m=10$ differed from the one for $m=11$
in the sixth significant digit.
Thus we conclude that
our energy calculations are correct to within one part in $10^5$
or so. To eliminate the danger that during a minimization process
we were accidentally
trapped in a local minimum, we repeated each run several times
starting with different initial configurations. An additional test is
provided by the accuracy with which we verify the virial relation
$$ W_b=2 W_e+W_v$$
obtained by the well known scaling argument of Derrick \refmark\rDe.
We define
$${\Delta}={{2 W_e+W_v-W_b}\over W_b} \eqn\ethreenine $$ and check
 that in
all our calculations $\Delta$ is smaller than $10^{-4}$. Finally, the
spherical symmetry of the ansatz provides another test of the accuracy
of our results. Notice that
the ansatz is spherically symmetric
in the sense that a spatial rotation by some angle $\beta$ can be
compensated by a global internal U(1)
rotation by $N\beta$. This means that the corresponding generator
 $Q-N L$ where $L$ is the
field angular momentum to be defined in the next section, must vanish
 for
the solution. $Q$ is zero by construction and we check, by explicit
evaluation of the integral of the angular momentum density, that for all
solutions $L$ is also zero to within one part in $10^4$.

Figure 2 shows the values of the ratio $R=E_{N=2}/{2 E_{N=1}}$
of the energy of the double vortex to twice the energy of the single
one,
as well as
the value of the energy of a single vortex itself devided by $\pi$
for various sets of
the parameters of the model.
For $\lambda=0$ and for static configurations our model reduces to
the Ginzburg-Landau equations for a superconductor and $\kappa$
corresponds exactly to the ratio of the penetration depth to the
coherence lenght defined there.

Figure 2d is in perfect agreement with the
results obtained previously in the context of the Ginzburg-Landau theory
\refmark\rA \refmark\rNO \refmark\rJR. We divided all energies by $\pi$
to make easier contact with previous results concerning
the Ginzburg-Landau theory, in which for $\kappa=1$ the $N=n$ vortex
solution has energy $E_n=n$.
The fact that the energy of a vortex
is an increasing function of the parameters $\kappa$ and $\lambda$
is easily
understood by the following argument: (we give the argument for
$\kappa$).
Start with the solution of the static
equations for some value of the parameter $\kappa$ and a given
$\lambda$ which
we keep fixed. By construction it satisfies Gauss' constraint and is
 neutral.
It can thus be used
as a trial initial configuration for the energy
minimization for a different value $\kappa^\prime < \kappa$. The energy
$E(\kappa^\prime)$ of the solution for $\kappa^\prime$ is smaller than
the value of the energy functional for the above configuration.
The latter
is equal to the energy of the solution for the value $\kappa$ minus the
positive quantity $(1/8)\;(\kappa^2-{\kappa^{\prime}}^2)\;
\int d^2x\;(\Psi
\Psi\sp\ast-1)\sp2 $, where $\Psi$ is the solution for $\kappa$.
Thus, $E(\kappa^\prime) < E(\kappa)$. Q.E.D. A similar
argument with a little extra care due to Gauss' constraint applies to
$\lambda$.
The other extreme, the no-potential case $\kappa=0$ is shown on figure
 2c.
The critical value of $\lambda$ for the $N=2$ vortex to have the same
energy
with two well-separated single vortices is 0.0167.

The critical line separating the parameter space into regions-I and -II,
according to the value of $R$ being smaller or larger than one
respectively,
is shown on figure 3. Regions-I and -II correspond in the
Ginzburg-Landau
case to ordinary type-I
and type-II superconductors.
It is interesting to compare the energy of a triple vortex to that of
three
single vortices far from each other.
In contrast to the Ginzburg-Landau case there exists inside region-I
another
critical line separating the parameter space into two subregions
according
to whether $E(N=3)$ is larger or smaller than 3 times $E(N=1)$.
 Consistent with
the well known facts about the Ginzburg-Landau model this line too
passes through the point $(\kappa=1, \lambda=0)$. The subregion with
$E(N=3)<3 E(N=1)$ is further split into two subsubregions depending
on the
value of the ratio $E(N=4)/4 E(N=1)$ being larger or smaller than one,
and so on.

In the context of a relativistic model we would talk about stable and
unstable higher-$N$ vortices. For instance, a vortex with $N=2$ and
energy
larger than twice the energy of a single one could not be
absolutely stable against decay to two single vortices. In the model
at hand though as we will see in the next chapter a completely
different picture arises \refmark\rPTb.

        \section{ The two-vortex system}

Our main goal in this section is to compute the energy
of two single vortices as a function of their separation $d$.
Again we proceed numerically
and choose a  variational ansatz
with the following characteristics: 1) The complex field
vanishes at the points $(\pm {d\over 2},0)$ on the x-axis. 2) For
large $d$ the
configuration reduces to two well-separated axially symmetric single
vortices at a distance $d$ from each other, while 3) for $d\to 0$ it
takes
the form of an axially symmetric $N=2$ vortex located at the origin.
For the phase $\Theta$ and the magnitude $X$ of the complex field $\Psi$
we write
\refmark\rJR
$$  \Theta (x,y) \;=\; tan \sp {-1}\bigl[
{{2x y\over x\sp{2} -y\sp{2}-{d\sp{2}\over 4}}} \bigr]    \eqn\etheta $$
$$X({\bf x}) \;=\; \omega \; X\sp{(1)}(\vert {\bf x} -
{{{\bf d}}\over 2}\vert)
\;  X\sp{(1)}(\vert {\bf x} \;+\; {{\bf d}\over 2}\vert)
+\;(1-\omega) \; {(\vert {\bf x} - {{{\bf d}}\over 2}
\vert)\; (\vert {\bf x} +{{{\bf d}}\over 2}\vert)\over r\sp2}\;
  X\sp{(2)}(r) \;+\;\delta X  \eqn\edxi $$
in terms of the previously determined solutions
$X\sp{(1)}$ and $X\sp{(2)}$
for the $N=1$ and $N=2$ vortices respectively. The function $\delta X$
is written in the form
$$ \delta X \;=\;(|{\bf x}-{{{\bf d}}\over 2}|)\; (|{\bf x} +
{{{\bf d}}\over 2}|)
  \;cosh\sp{-1} (k r) \;
  \sum_{m=0}\sp{l}\sum_{n=0}\sp{m}\;z_{mn}\; r\sp{2m}\;cos(2n\phi)
  \eqn\evdxi $$
$ \omega $ is a variational parameter, the relative weight of
$X\sp{(1)}$ and $X\sp{(2)}$ in the configuration. To fix its optimal
value we minimized
   the  energy with respect to $ \omega $ for  $\delta X=0 $.

The variational ansatz is completed by a trial set of functions for
the gauge
potential
 $$ {\bf A}({\bf x}) \;=\; \omega \; {\bf A}\sp{(1)}
  (\vert {\bf x} - {{{\bf d}}\over 2}
\vert)\;+\; \omega \;
  {\bf A}\sp{(1)}(\vert {\bf x} \;+\; {{\bf d}\over 2}\vert)
\;+\;(1-\omega) \;
  {\bf A}\sp{(2)}(r) \;+\;\delta {\bf A}({\bf x})
  \eqn\edalp $$

where $\delta {\bf A} \;=\;\delta A_r\;\hat r\;+\;\delta A_
  {\phi}\;{\hat \phi} $ with $\delta A_r$ and $\delta A_{\phi}$
  defined by
   $$  \delta A_r\;=\;r\;
  cosh\sp{-1} (r) \;
  \sum_{m=0}\sp{l}\sum_{n=0}\sp{m}\;\alpha_{mn}\sp{r}
  \;r\sp{2m}\;sin(2n\phi)
  \eqn\evdalp $$
   $$  \delta A_{\phi}\;=\;
 r\; cosh\sp{-1} (r) \;
 \sum_{m=0}\sp{l}\sum_{n=0}\sp{m}\;\alpha_{mn}\sp{\phi}
 \;r\sp{2m}\;cos(2n\phi)
  \eqn\evdalp $$

As mentioned above ${\bf d}=(d,0)$ to place the two vortices
at $(\pm {d\over 2},0)$
along the $x$ axis. The angular dependence of the fields
was restricted to the form given above by the requirement on
the configuration
to be invariant under spatial reflection
with respect to either one of the coordinate axes.

We used the same numerical procedure as in the isolated vortex case
to determine the optimum values of the remaining
variational parameters $\alpha_{ij}\sp{\phi},\alpha_{ij}\sp{r}$,
$z_{ij}$ and $k$.
The dominant source of error in our results is due to the truncation
of the
configuration space. Again an estimate is obtained from the change in
the
value of the quantity of interest as we vary the number of variational
parameters. To achieve an error of the order of $0.1\%$ in the total
energy
for all values of $d$, it was enough to set $l=1$ for $d>3$, while
for $d<3$
it was necessary to take $l=2$  i.e. a richer variational ansatz with
18 parameters. The uncertainty in
$W_e$ was a little larger, something like $0.2-0.3\%$, but since $W_e$
contributes always a small fraction to the total energy, this error is
negligible. Contrary to the isolated vortex case we did not perform the
spatial integrations analytically. Instead we carried out these
integrations
numerically. Appropriate choice of the grid and the boundaries reduced
the corresponding error to the order of $0.01\%$.

In figure 4 we plot the energy divided by $\pi$ of the two-vortex
system as a function of their separation $d$ for four sets of the
parameters $\kappa$ and $\lambda$
corresponding to the points $A,B,C$ and $D$ shown in figure 3,
from well inside region-II, to somewhere
in region-I. We have included the point $C$ on the critical line with
$(\kappa=0.5\;,\;\lambda=0.0115)$.
Figure 4a corresponds to parameter values far from the critical line
in region-II. The force
between the vortices is everywhere repulsive.
In all other cases one immediately recognizes a region of attraction
and a
region of repulsion of the two solitons. The local extremum in the
energy
leads to the exciting possibility of the existence of dumb-bell shaped
doubly-charged static solutions of the field equations for a
rather wide range of parameters. We will have the opportunity to
elaborate on this in a future
publication \refmark\rST.

Finally, figures 5, 6, 7 and 8 show the total energy
density $w_t$, the gauge-invariant
topological density $\tau$, the magnetic field $B$ and the
electrostatic energy distribution $w_e$ respectively,
of the optimum two-vortex configuration
corresponding to the parameter values $(\kappa=0.5\;,\;\lambda=0.0115)$
on the critical line. The pictures in each one of them refer to
distances $d=0,\;2,\;4,\;6$. Notice that of the two definitions
$B/{2 \pi}$ and $\tau$
of the topological density the former follows less closely the
energy distribution and as such it offers a bad description
of the two vortex positions. We have checked, although we do
not show it here,
that the topological density
in terms of the complex field alone is even worse in that respect.
Notice also that the distribution of $w_e$ does not
follow that of the total energy.
It spreads over the whole region between the two vortices
and is actually responsible for the bump in their interaction
energy. Being so small
in magnitude it does not alter the final picture that the two vortices
can be safely considered far from each other
already for $d=4$.
These features of $w_e$ as well as its negligible magnitude compared to
$w_t$ are also shown on figure 9 for the case of an axially symmetric
$N=1$ configuration with $\kappa=0.5$ and $\lambda=0.0115$. The virial
relation given in the previous section shows that $W_e$ is
never dominant. It
can at most be $1/3$ of $W_t-W_d$ for any value of the parameters.

We end this section with a remark about the numerical procedure used
to compute $W_e$ at each step of the iterative minimization process.
We exploited the linearity of the Poisson equation to compute
once and for all
the electric field created by the charge density due to the
configuration (3.10) to (3.12) in terms of the parameters of the
ansatz. Thus the calculation of the electric field after each
iteration is immediate. One does not have to solve the Poisson
equation again but
just plug into the general formula the new values of the
parameters. $W_e$ is finally given by a straightforward spatial
integration.
This way the computation is more accurate and much faster.

\chapter{Vortex Dynamics}
\section{Canonical structure of the model}

In order to investigate the dynamics of vortices we start with
the canonical structure
of the model. We follow the standard procedure \refmark\rL applied to
gauge
theories to determine its fundamental Poisson brackets.
As mentioned in a previous section to eliminate the redundant degrees
of freedom
due to the gauge invariance of the model we impose the condition
$$\nabla \cdot \bf A=0$$

Since the action does not depend on $\dot{A_0}$, $A_0$ is a dependent
variable
satisfying the Gauss law constraint which we solve to obtain
$$ A_0({\bf x},t)=-{\lambda\over {2\pi}}\; \int d^{2}x^{\prime}\;
ln|{\bf {x-x^{\prime}}}|\;
(\Psi^{\ast} \Psi({\bf x^{\prime}},t) - 1) \eqn\efourone $$
(use was made of $\nabla^{2}\;ln|{\bf {x-x^{\prime}}}|=2 \pi \delta
({\bf{x-x^{\prime}}})) $. With the above gauge-fixing condition
the purely electromagnetic part of the action takes the form
$$ S_A={1\over 2} \int d^{2}x\;[\;{1\over \lambda} ({{\bf E}_T}^2 +
{{\bf E}_L}^2 )- B^2\;]
\eqn\efourtwo $$
where the transverse and the longitudinal parts of the electric field
 are
${\bf E}_T=-\dot{\bf A}$ and ${\bf E}_L=-\nabla A_0$ respectively.

The canonical momentum $\pi_k$ conjugate to $A_k$ is then $\pi_k=
\partial {\cal L}/\partial {{\dot A}_k}=-{E^T_k}/\lambda$, while $\Psi$
 and
$\pi\equiv \partial {\cal L}/\partial {\dot \Psi}
=i\Psi^{\ast}$ are conjugate of each other. Thus, the fundamental
Poisson brackets are
$$ \{\Psi({\bf x},t), {\Psi^{\ast}}({\bf x}^\prime, t)\}=
-i\delta(\bf x-x^\prime)$$
$$ \{A_i({\bf x},t), {\dot A}_j({\bf x}^\prime,t)\}=\lambda
{\delta^T}_{ij}(\bf x-x^\prime) \eqn\efourthree $$
$$ \{\Psi({\bf x},t), {A_i}({\bf x^\prime},t)\}=0=
   \{\Psi({\bf x},t), {{\dot A}_i}({\bf x^\prime},t)\}$$
where the transverse $\delta$ function is defined as
${\delta^T}_{ij}(\bf x-x^\prime)\equiv (\delta_{ij}-\nabla^{-2}
\partial_i \partial_j)\; \delta(\bf x-x^\prime)$.
Finally, the Hamiltonian (energy) of
the system is
$$ H=\int d^2x \;[\;{1\over {2 \lambda}} ({{\bf E}_T}^2 +
{{\bf E}_L}^2 )+ {1\over 2} B^2 + {1\over 2} |{\bf D}\Psi|^2 +
{1\over 8} \kappa^2 (\Psi^\ast \Psi - 1)^2\;]
\eqn\efourfour $$

All other Poisson brackets are obtained from the ones given above.
It is
straightforward to check that
$$ \{A_0({\bf x},t), \Psi ({\bf x^\prime},t)\}=-{{i\lambda}\over
{2 \pi}}
ln|{\bf x-x^\prime}| \;\Psi ({\bf x^\prime},t)$$
$$ \{A_0({\bf x},t), {A_i} ({\bf x^\prime},t)\}=0=\{A_0({\bf x},t),
{{\dot A}_i}({\bf x^\prime},t)\} \eqn\efourfive $$
as well as that the equations of motion \etwoseven$\;$  and \etwoeight
$\;$coincide with
Hamilton's equations
$$ {\dot \Psi}({\bf x},t)=\{ \Psi ({\bf x},t), H\}$$
$$ {{\dot A}_k}({\bf x},t)=\{{A_k}({\bf x},t), H\} \eqn\efoursix$$
$$ {{\dot \pi}_k}({\bf x},t)=\{{\pi_k}({\bf x},t), H\}$$

We are now ready to proceed with the construction of the field momentum
and angular momentum for the model at hand, valid in all topological
sectors. We will show \refmark\rPTb \refmark\rPTa that the
corresponding conservation laws describe the basic feature of vortex
motion in the absence of external forces, namely spontaneous
pinning.

\section{Momentum and angular momentum in all sectors}

The Noether expression for the linear momentum in our model is
$$ P^N_k=\int d^2x \;(-\pi {\partial_k} \Psi - {\pi_j} {\partial_k} A_j)
\eqn\efourseven $$
It is derived under the assumption that all fields approach their
vacuum values fast enough as $r\to \infty$ so that all integrals
are meaningfull and surface terms
appearing in intermediate steps are zero. It is straightforward
to check that the above
expression is ill-defined for any vortex configuration with non-zero
topological charge. Under the same assumptions though which led to
\efourseven $\;$ the latter can be brought to the form
$$ P_k=-\epsilon_{ki} \int d^2x \;x_i \;\epsilon_{lm} (\partial_l \pi\;
\partial_m \Psi + \partial_l \pi_j \;\partial_m A_j) \eqn\efoureight $$
which is well-defined for any smooth finite-energy field
configuration with arbitrary topological charge. The two expressions
$\efourseven$ and $\efoureight$
differ by exactly those surface terms which were omitted from the
former and make the latter finite and unambiguous in all topological
sectors.

All the defining properties of the momentum are
verified for $P_k$. This is no surprise.
They are formally valid for $P_k^N$ and this differs from $P_k$
only by surface terms.
In any case, one can directly and unambiguously show
using the equations of motion $\etwoseven$, $\etwoeight$ and the Poisson
brackets derived above
first, that $P_k$ is conserved
$${d\over dt} P_k=0 \eqn\efournine $$
second, that it is the generator of spatial displacements
$$ \{ P_k, {\cal F}\}=\partial_k {\cal F} \eqn\efourten $$
and finally, that it is gauge-invariant. The only condition on the
configuration
${\cal F} \;=\;(\Psi,A_0,A_i)$ necessary for the derivation is
that $A_0$ vanishes faster than $r^{-1} \;$ as $r\to \infty$. All our
vortices have $A_0$ approaching zero at infinity exponentially
fast and consequently they safely belong to this set of
configurations. Furthermore,
using Gauss' constraint we can rewrite the momentum in the manifestly
gauge-invariant form \refmark\rPTb
$$ P_k=\epsilon_{ki} \int d^2x\; (2 \pi\; x_i\; \tau + {1\over
\lambda}\;E_i B)
\eqn\efoureleven $$
The second term is the Poynting vector, the pure
electromagnetic contribution to the field momentum.

Thus, independently of derivation, $P_k$ is the correct form of the
momentum in our theory, valid in any topological sector and
reducing to the naive expression in the trivial sector $N=0$
and for configurations
allowing for free integrations by parts.

Before we discuss the implications of the above formula of the
linear momentum
on the motion of vortices, we would like to construct in a similar
way the
correct form of the angular momentum in our model.
Again, the Noether expression
$$ L_N=- \int d^2x\; \epsilon_{ij} \;[\;x_i\;(\pi \partial_j \Psi +
\pi_k \partial_j A_k) + \pi_i A_j\;] \eqn\efourtwelve $$
is formally conserved and generates rotations but is divergent in
any non-trivial sector. Allow for integrations by parts and ignore
surface terms to
rewrite it in the form
$$L=\int d^2x\;[\;{1\over 2} {\bf x}^2 \; {\epsilon}_{ij}\;
(\partial_i \pi \partial_j \Psi + \partial_i \pi_k \partial_j A_k) +
{\epsilon}_{ij} A_i \pi_j\;] \eqn\efourthirteen $$
or as a manifestly gauge invariant quantity in terms of the
topological density $\tau$ defined in section 2
$$L=-\int d^2x\; ( \pi {\bf x}^2 \tau + {1\over \lambda}\;
{\bf x}\cdot {\bf E} B)
\eqn\efourfourteen $$

Using either one of the last two formulas one can show that for
essentially
all finite-energy configurations as in the case of the linear momentum,
$L$ is well-defined, conserved and generates spatial rotations, i.e. it
satisfies
$$\{L,\Psi({\bf x},t)\}={\epsilon}_{ij} x_i \partial_j \Psi$$
$$\{L,{A_k}({\bf x},t)\}={\epsilon}_{ij} x_i \partial_j A_k +
{\epsilon}_{kl}
 A_l \eqn\efourfifteen $$

$L$ is then the proper definition of the angular momentum when dealing
with topologically non-trivial configurations. As already explained
it vanishes for the spherically symmetric neutral
solutions studied in section three.

\section{Vortex motion}

In a following paper $\refmark\rST$ we will present
details of a numerical simulation
of the motion of the flux vortices described above under the influence
of various external forces.
Here we would like for the sake
of completeness to predict analytically and without any
approximation the essential features of their dynamical behaviour
\refmark\rPTb.
It is already apparent that
the momentum $P_k$
defined in the previous section is not a measure of the translational
motion of a vortex but instead it describes its position. In fact the
momentum of a static axially symmetric vortex with charge $N$
centered at $\bf a$ is $P_k=2\pi N \epsilon_{ki} a_i$.
It is also clear that a localized free vortex of arbitrary shape
moving in
formation at constant velocity $v_i$ would have $P_k=2
\pi N \epsilon_{ki}
(a_i^0+v_i t)$. For any $v_i\ne 0$ this is forbidden by the linear
momentum
conservation law. A free vortex is spontaneously pinned as a consequence
of momentum conservation.

Let us define the "guiding center" $\bf R$ of a generic configuration
with $N\ne 0$ by
$$  R_i\equiv -{1\over 2\pi N} \epsilon_{ij} P_j \eqn\efoursixteen $$
Since under a rigid displacement of the whole configuration by $\bf c$,
$\bf R$
changes to $\bf R + c$ and for a spherical vortex it coincides with
its geometric center, we naturally interpret it as the mean position
of the configuration. Note also in support of this interpretation, that
$\bf R$ is related to the first moment of the topological density
$\tau$,
so that for nearly spherical configurations $\bf R$ is close
to their "center of topology".

{\bf a.} In the absence of external forces both $N$ and $\bf P$ are
conserved and
$${d\over dt} \bf R=0 \eqn\efourseventeen $$
A generic vortex-like configuration produced in the system
will of course
fluctuate in its details but it will remain pinned
at its initial mean position.

{\bf b.} Consider next the response of such a vortex to an externally
prescribed electric current ${\tilde J}_i ({\bf x}, t)$. Its effect is
studied by adding to the action the term $\delta S= \int d^2x \;dt\;
{\tilde J}_i ({\bf x}, t) A_i$. For concistency of the model the
external
current should have zero divergence. The new term in the action
modifies the $A_i$ equation of motion $\etwoseven\;$ by the substitution
$J_i\to J_i+{\tilde J}_i$ on the right-hand side. Because of the
external
current the momentum and the angular momentum are no longer
conserved. Instead, their time derivatives
are equal to the external force and torque respectively. Indeed one can
easily verify
Newton's equation:
$${d\over dt} P_k= F^{Lorentz}_k=-\int d^2x \; \epsilon_{kl}
{\tilde J}_l ({\bf x}, t) B({\bf x}, t) \eqn\efoureighteen $$
as well as
$${d\over dt} L=Torque=\int d^2x \;x_i \tilde J_i B \eqn\efournineteen $$
Naturally the force on the vortex is opposite to the Lorentz force acting
on the external current, while the torque is by definition expressed
in terms of the
force density $f_k=-\epsilon_{kl} \tilde J_l B$ as the integral of
$\epsilon_{ij} x_i f_j$ . We now use $\efoursixteen$ and the fact
that for arbitrary $\tilde J$ $N$ is still conserved to obtain
$$ {dR_k\over dt}=-{1\over 2 \pi N} \int d^2x \;{\tilde J}_k B
\eqn\efournineteen $$

In the idealized situation of a homogeneous
throughout the vortex external current the above formula simplifies
to
$$ {dR_k\over dt}=-{\tilde J}_k (t) \eqn\efourtwenty $$

In general, the naive expectation based on the usual Newtonian
reasoning and Galilean invariance would be that
the vortex should accelerate in the direction
of the force acting on it. Instead, $\efoursixteen$ combined with
${dP_k}/dt=F_k$ for a generic force $F_k$, shows that
the equation of motion of the mean position of the vortex is
$$ {d\over dt} R_k=-{1\over {2 \pi N}} \;\epsilon_{kl} \;F_l
\eqn\efourtwentytwo $$
i.e. the vortex moves with speed ${\bf |F|}/{2 \pi |N|}$ and at
$\pm\; 90^0$ relative to the force for positive or negative $N$
respectively.
In the special case of the homogeneous external current considered here,
the force is also proportional to $N$ and all vortices move with the
same velocity equal to minus the external current itself.
If in particular the latter
is due entirely to the condensate charges taken here by convention
positive with unit charge density, its value
is exactly equal to minus the velocity of the carriers.
Thus the vortex will reorganize itself during
a transient period following
the onset of the external current and it will move with the same
speed but opposite to
the current carriers.
Furthermore it should
be pointed-out that in the context of our field theory model no
approximation other than the implicit
assumption that the vortex
remains localized was necessary. The position interpretation
$\efoursixteen$ of
the momentum converts Newton's law into an equation giving directly
the velocity of the vortex in terms of the applied force, while
the $\epsilon_{ij}$ of $\efoursixteen$ makes
the vortex move at an angle $(N/|N|)\;90^0$ relative to the direction
 of
the external force (Hall-behaviour) $\refmark\rPTb \refmark\rPTa$.
The above general conclusion, reached without ever solving an
equation of motion, does not depend on the details
of the Hamiltonian of the system.
Any potential $V$ or any additional higher spatial
derivative terms in $H$ modifies the detailed profile of the
vortex solutions but it does not alter their dynamics.

The guiding center interpretation of the momentum $\efoursixteen$ is not
applicable in the case of topologically trivial $(N=0)$ configurations.
Newton's
law $\efoureighteen$ is of course still true for any external
force, but this does not strictly speaking tell us much about
the actual motion of the vortex. In contrast to the case of ordinary
particle dynamics with Poincar$\acute e$ or even Galilean
invariance the relation between momentum and velocity is
not a priori known in the present model.

\chapter{Discussion}

We presented a general treatment of the motion of flux-vortices
in a two parameter field theory model describing as stated in the
introduction
a rather wide class of idealized physical systems.
We argued that
independently of the details of the Hamiltonian the vortices exhibit
Hall-behaviour, like electrons moving on a plane under the simultaneous
action of a perpendicular magnetic and an in-plane electric field.
The picture
is identical to the one derived previously for the dynamics of magnetic
bubbles in the ferromagnetic continuum \refmark\rPTa. The fact that the
solitons in these two field theories are topologically different,
classified by the second homotopy group of $S^2$ in one case
and by the first homotopy group of $S^1$ in the other, does not
seem to play any role in our discussion.

An external
current was shown from first principles
to pull the vortex in the opposite direction. It is not clear at
this point what is the relation of our model with the ones
considered previously \refmark\rH \refmark\rNV
in connection with superconductivity. It is tempting to think of it as
a field-
theoretic realization of the hydrodynamic model discussed
in reference [16] in the vanishing friction limit.
The conclusion though reached there about vortex motion,
based on the phenomenological use of the Magnus force
\refmark\rFGM known from
fluid mechanics \refmark\rB,
is $\bf V=\tilde J$. This is opposite to ours, which is correspondingly
reminiscent of the opposite sign Hall effect reported in high-$T_c$
superconductors $\refmark\rDo$.

The Hall
motion of the vortices derived above is attributed
to the radical change in the
role of the momentum of the theory as a result of the underlying
topology. The Hall behaviour of an isolated vortex is exactly
due to the fact that the linear momentum $\efoureleven$
contains a piece
which is equal to the first moment of the topological density.
An immediate consequence of this fact combined with the fundamental
property
$\efourten$ of the momentum is that
$$\{P_1,P_2\}=2 \pi N \eqn\efiveone $$

We thus conclude that necessary condition for the manifestation of Hall
motion is that the translation part of the symmetry algebra of
the model admits central extension. It is well-known $\refmark\rLeLe$
that this is not possible in the case of the Galilean or the
Poincar$\acute e$ algebra in any spatial dimension higher than or equal
to two.
It is not even true for the Euclidean algebra $E(D)$ for $D\ge 3$. Thus,
the Hall behaviour encountered above is expected a priori only in two-
dimensional systems with spatial $E(2)$ algebra, or in $D$-dimensional
systems with translational symmetry alone.
In the system at hand one can immediately check that
$$\{L,P_1\}=P_2 \hskip 2cm \{L,P_2\}=-P_1 \eqn\efivetwo $$
and verify that it belongs to the first category as expected.

We would also like to stress that although $\efoureight$ is
true even in theories with canonical momentum $\pi$
proportional to the time derivative
of the complex field, our previous reasoning does not go through
since $\bf P$ is then related to the first moment not of the topological
density but of a quantity which actually vanishes for static
configurations. Thus,
another condition which is necessary is that the equations
of motion are first order in the fields which carry the topology
in the particular model. In other words, we do not expect Hall-
motion of the vortices in the relativistic-like model $\refmark\rW$
derived as a dynamical extension of the
Ginzburg-Landau theory of a superconductor appropriate for temperatures
far below $T_c$.

On the other hand, the above formalism applies to the motion of
vortices in a generalization of the present model which allows for
dynamical positive-ions forming a lattice, of
the $(N_h,N_{ss})$ with $N_{ss}\ne -2 N_h$ vortices
of the high-$T_c$ superconductor model proposed in reference [21],
as well as to the motion of vortices in easy-plane
ferromagnetic films $\refmark\rMBWK$, to mention just a few systems
of considerable interest.

\ACK\
We would like to thank N. Papanicolaou for helpful discussions.
This work was supported in part by the EEC grant No SC1-0430-C and by
the Greek General Secretariat of Research and Technology grant
No $91E\Delta
358$.

\vfill\eject

\vskip 2cm
\hskip 0.5cm {FIGURE CAPTIONS}

Figure 1: $N=1$ vortex profiles. (1a) and (1b) correspond to $\kappa=2$
and $\lambda=0.1$, while (1c) and (1d) to $\kappa=0$ and $\lambda=0.1$

Figure 2: Plots of $R=E_{N=2}/2 E_{N=1}$ and of $E_{N=1}/\pi$ as
functions
of $\kappa$ or $\lambda$ for the values of the second parameter as shown.
(2d) corresponds to the Ginzburg-Landau model, while (2c) to the no-
potential case.

Figure 3: Regions-I and -II of the parameter space.

Figure 4: The energy of the two-vortex system as a function of their
separation for the values of parameters corresponding to the points
$A,\;B,\;C$ and $D$ of figure 3. The dashed line is drawn at twice the
energy of a single vortex.

Figure 5: The total energy density of the two vortices
$d=0,\;2,\;4$ and $6$. $\kappa=0.5$ and $\lambda=0.0115$.

Figure 6: The topological charge density $\tau$ of the two
vortices for
$d=0,\;2,\;4$ and $6$. $\kappa=0.5$ and $\lambda=0.0115$.

Figure 7: The magnetic field of the two vortices for
$d=0,\;2,\;4$ and $6$. $\kappa=0.5$ and $\lambda=0.0115$.

Figure 8: The electrostatic energy density of the two vortices
$d=0,\;2,\;4$ and $6$. $\kappa=0.5$ and $\lambda=0.0115$.

Figure 9: Plot of the densities $w_t$ and $150 w_e$ for an
isolated $N=1$ vortex.
$\kappa=0.5$ and $\lambda=0.0115$.

\vfill\eject

\refout

\bye
\end